\documentclass[]{aastex62}

\accepted{for publication in RNAAS }
\begin{document}

\title{Discovery of a magnetic white dwarf with unusual short-period variability}

\author[0000-0001-8993-5053]{Aleks Scholz}
\affiliation{SUPA, School of Physics \& Astronomy, University of St Andrews, North Haugh, St Andrews, KY16 9SS, United Kingdom}

\correspondingauthor{Aleks Scholz}
\email{as110@st-andrews.ac.uk}

\author[0000-0003-4450-0368]{Joe Llama}
\affiliation{Lowell Observatory, 1400 W. Mars Hill Rd. Flagstaff, AZ 86001, USA}

\author{Koraljka Muzic}
\affiliation{SIM/CENTRA, Faculdade de Ciencias de Universidade de Lisboa, Ed. C8, Campo Grande, P-1749-016 Lisboa, Portugal}

\author{Sarah Faller}
\affiliation{SUPA, School of Physics \& Astronomy, University of St Andrews, North Haugh, St Andrews, KY16 9SS, United Kingdom}

\author{Dirk Froebrich}
\affiliation{Centre for Astrophysics and Planetary Science, University of Kent, Canterbury, CT2 7NH, UK}

\author{Beate Stelzer}
\affiliation{Institut f{\"u}r Astronomie und Astrophysik, Eberhard Karls Universit{\"a}t, Sand 1, D-72076 T{\"u}bingen, Germany}
\affiliation{INAF-Osservatorio Astronomico di Palermo, Piazza del Parlamento 1, 90134, Palermo, Italy}


\section*{}

We report the discovery of a magnetic white dwarf which shows periodic variability with $P=110$\,min, color-dependent amplitudes and a transient phase shift in the blue compared to the red lightcurve -- a previously unknown type of variability for this type of object. We attribute the variations either to a close ultracool (thus far undetected) companion or, more likely, to magnetic spots with unusual temperature structure.

The object SDSS 160357.93+140929.97 (hereafter SDSS\,16+14) was identified as a magnetic white dwarf in \citet{2013MNRAS.429.2934K}, with an effective temperature of 10123\,K and a magnetic field strength of 43\,MG. We monitored this target on May 3rd and June 5th 2015, each time over a time period of 3\,h, using the optical camera EFOSC2 at the ESO/NTT (program 095.D-0245). The g-band lightcurves shows a photometric period of $\sim 100$\,min, with a min-max amplitude of $\sim 3$\% (compared to a photometric noise of 1-2\%).\footnote{As part of the same run we also confirm in broadband light the periods for SDSS\,151415.66+074446.50 and SDSS\,125044.43+154957.36, both known to have short-period RV companions \citep{2012MNRAS.423.1437B}.}

\begin{figure}
\begin{center}
\includegraphics[scale=0.7,angle=0]{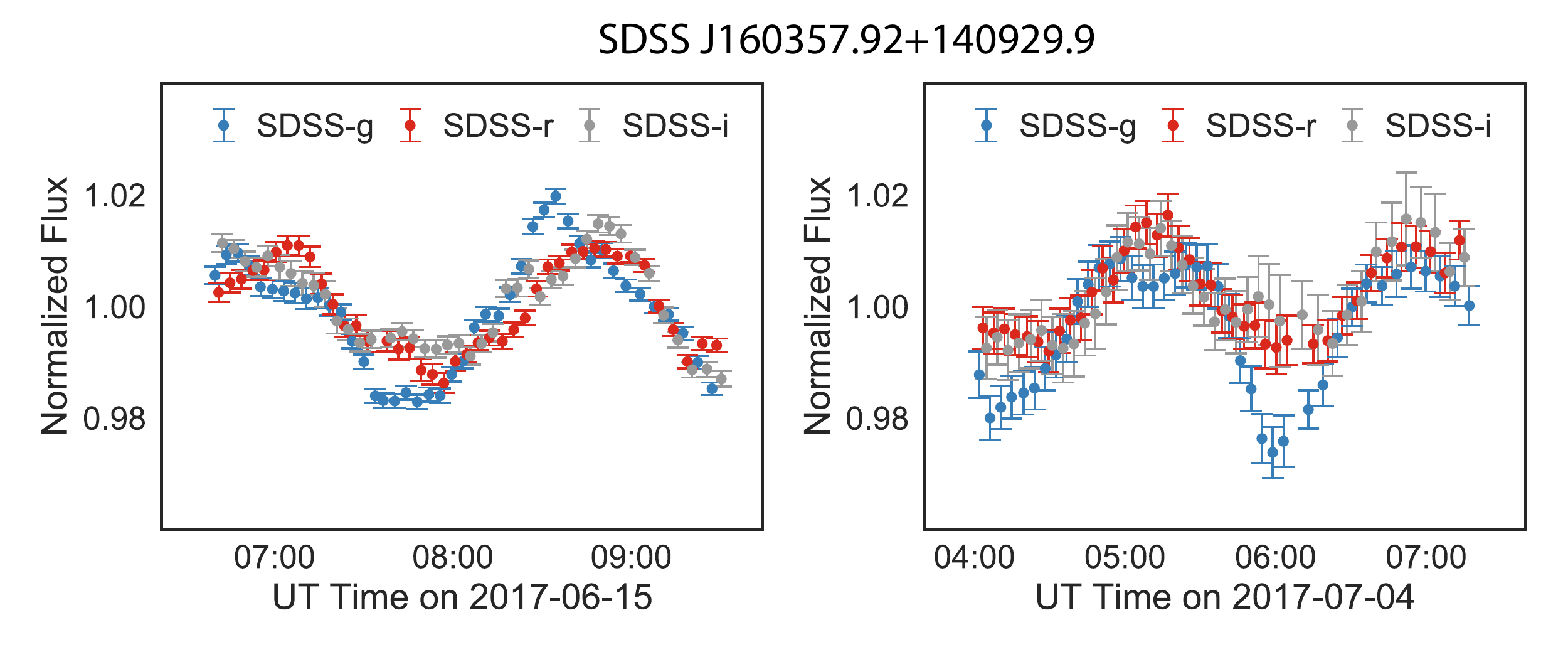}
\caption{DCT/LMI lightcurves in multiple bands for SDSS\,16+14.
\label{fig}}
\end{center}
\end{figure}

We obtained high signal-to-noise lightcurves for SDSS\,16+14 using Lowell Observatory's Discovery Channel Telescope (DCT) and the Large Monolithic Imager (LMI), on June 15th and July 4th 2017. Here we observed in the three filters g, r, i quasi-simultaneously, i.e. switching filters every few minutes. The resulting multi-band lightcurves are shown in Figure \ref{fig}. This dataset confirms the presence of a photometric period with $P = 110 \pm 3$\,min, averaged over the six lightcurves. The amplitudes in g, r, and i are 0.027, 0.021, 0.021 in the first run and 0.030, 0.021, 0.019 in the second run, respectively. Interestingly, the first set of observations shows a clear offset in phase of $\sim 0.2$ between the r/i-band lightcurves and the one in the g-band. This offset was not observed (or is significantly smaller) in the second run. Both times, the g-band lightcurve is deviating from the sinusoidal variations seen in r- and i-band.

There are two ways to explain this type of variability. The first option is to postulate the presence of a cool, unresolved companion, with the period being the orbital period. The observed period and amplitude fit into the small group of white plus brown dwarf systems in the literature \citep{2018arXiv180107773C}. The variability could be produced by reflected light, thermal re-radiation, accretion, or beaming \citep{2015MNRAS.447.1749M}. The phaseshift may be caused by more than one of these mechanisms being involved. To validate this hypothesis, we searched for the IR excess caused by the companion, using deep H- and K-band imaging from Subaru/IRCS, in an observing run on July 1-2 2017. The object is clearly detected as a single point source, and we measure magnitudes of $H=18.1 \pm 0.1$ and $K=18.3\pm 0.1$. Within the errorbars, this is consistent with a 10000\,K blackbody scaled to the Sloan g-band magnitude of SDSS\,16+14 and rules out L-type companions with $T>1000$\,K. Radial velocity monitoring is needed to exclude even cooler companions.

In absence of an IR excess, the favoured explanation is the presence of magnetic spots and a rotational modulation of the flux. In terms of period, amplitude, temperature and magnetic field strength, SDSS\,16+14 fits into the sample of magnetic white dwarfs with measured rotation periods by \citet{2013ApJ...773...47B}. Simple blackbody spot models, such as used by \citet{2009MNRAS.398..873S}, match the color dependence of the amplitudes, with spot temperature around 5000\,K and a filling factor of 2-3\%. If this interpretation is correct, SDSS\,16+14 is one of the fastest known rotating white dwarfs \citep{2015ASPC..493...65K} and among the spotted white dwarfs with the strongest magnetic field. The phaseshift observed in one of the DCT runs requires an unusual temperature structure in the spots (e.g., a hot spot and a cool spot offset in longitude). This finding encourages multi-band observations of variable magnetic white dwarfs to elucidate whether or not such spot configurations are common.

\acknowledgments

We thank Sarah Casewell, Elme Breedt, Philip Lucas, Hiro Takami, and Brendan McDonald for supporting this project. These results made use of Lowell Observatory's Discovery Channel Telescope, and The Large Monolithic 
Imager (NSF/AST-1005313).

\facilities{DCT,NTT,Subaru}

\end{document}